\documentstyle[12pt]{article}
\newcommand{\m}{\medbreak}

\newcommand{\no}{\noindent}
\newcommand{\EQ}{\begin{equation}}
\newcommand{\eq}{\end{equation}}
\newcommand{\EQA}{\begin{eqnarray}}
\newcommand{\eqa}{\end{eqnarray}}
\newcommand{\ds}{\displaystyle}
\newcommand{\AR}{\renewcommand {\arraystretch}{1.5}
\begin{array}{l}}
\newcommand{\bAR}{\renewcommand {\arraystretch}{2}
\begin{array}{l}}
\newcommand{\ARc}{\renewcommand {\arraystretch}{1.5}
\begin{array}{c}}
\newcommand{\bARc}{\renewcommand {\arraystretch}{2}
\begin{array}{c}}
\newcommand{\ar}{\end{array} \renewcommand {\arraystretch}{1}}

\newcommand{\ST}{\renewcommand {\arraystretch}{2}}
\newcommand{\st}{\renewcommand {\arraystretch}{1}}
\newcommand{\ee}{\mbox{$e^+e^-\ $}}

\newcommand{\ET}{\mbox{$E_T\ $}}

\newcommand{\ALLPV}{\mbox{$A_{LL}^{PV}\ $}}

\newcommand{\r}{\rightarrow}

\newcommand{\Z}{$Z^{\circ}\ $}
\newcommand{\ZP}{$Z'\ $}

\begin{document}
\begin{titlepage}
\vspace{0.2in}
\vspace*{1.5cm}
\begin{center}
{\large \bf Spin asymmetries
in one-jet production at RHIC with polarized proton beams :
the effects of a hadrophilic \ZP 
\\} 
\vspace*{0.8cm}
{\bf P. Taxil} and {\bf J.M. Virey}{$^1$}  \\ \vspace*{1cm}
Centre de Physique Th\'eorique$^{\ast}$, C.N.R.S. - Luminy,
Case 907\\
F-13288 Marseille Cedex 9, France\\ \vspace*{0.2cm}
and \\ \vspace*{0.2cm}
Universit\'e de Provence, Marseille, France\\
\vspace*{1.8cm}
{\bf Abstract \\}
\end{center}
We show that the measurement of some parity violating asymmetry in the
production of a large \ET jet could reveal the presence of a new
hadrophilic \ZP such
as the one recently introduced to interpret possible departures from the
Standard
Model predictions both at LEP and at CDF. Such a measurement could be
performed
within a few years by the RHIC Spin Collaboration (RSC) using the
Relativistic
Heavy Ion Collider (RHIC) as a polarized proton-proton collider.

\vfill
\begin{flushleft}
\m\no
Number of figures : 2\\

\m\no
April 1996\\
CPT-96/P.3333\\
\m\no
anonymous ftp or gopher : cpt.univ-mrs.fr

------------------------------------\\
$^{\ast}$Unit\'e Propre de Recherche 7061

{$^1$} Moniteur CIES and allocataire MESR \\
email : Taxil@cpt.univ-mrs.fr
\end{flushleft}
\end{titlepage}

\section{Introduction}
\indent
\m
Very recently, the CDF collaboration has reported an excess of jets
at large \ET with respect to the QCD prediction \cite{CDFa}. 
Although these
anomalies have to be confirmed, they have triggered new speculations
about the possible existence of New Physics beyond the Standard Model
(SM) whose direct effects could be within the reach of forthcoming
experiments.

The most straightforward interpretation of these data is to invoke
the presence of a quark substructure. If the effect on the high \ET jet
spectrum is parametrized conventionally \`a la Eichten et al.
\cite{EichtenEHLQ},
the value obtained by CDF for the compositeness scale $\Lambda$ is
$\Lambda\, =\, 1.6$ TeV.

On the other hand, some authors \cite{A,C} have related
these anomalies to the LEP results on $R_b$ and $R_c$ which deviate from
the SM predictions by about 3.5$\sigma$ and 2.$\sigma$ respectively
\cite{LEPcollab}. The idea is to introduce an extra heavy neutral vector
boson
(called \ZP in the following) which couples dominantly to quarks. Then a
small
\Z - \ZP mixing can modify the rate of $b-\bar b$ and $c-\bar c$ pair
productions
at the \Z peak provide that the direct \ZP couplings to quarks are
sufficiently
large. In the same time, in hadronic collider experiments, this \ZP, with a
mass
around  0.8 - 1.0 TeV and large couplings to quarks,
can mediate a new interaction between quarks via virtual
exchange , hence affecting the probability of
producing energetic jets.

Since these papers have appeared, it has been stressed that,
in the production of a top quark pair at FNAL, the additional contribution
due to
this \ZP improves the agreement between experiment and theory \cite{GStop}.
A more
precise measurement of $\sigma_{t \bar t}$ should clarify this point.
Concerning \ee physics,  the potentialities of LEP2 for the observability 
of hadrophilic \ZP effects have been explored in Ref.\cite{C}.

It is of some importance to investigate if some other kind of decisive
signal
could be obtained within a few years, using machines and detectors which
are now
under construction.
In fact, a new neutral vector boson, whose couplings to quarks exhibit a
particular chiral structure, will be at the origin of some new parity
violating
(PV) effects in many processes.  Such effects have been
investigated some time ago in the context of LEP2 \cite{Blondel}
and of the NLC \cite{NLC}(with polarized beams in both cases). 
In this respect, hard scattering in hadronic processes present a particular
interest 
-provide that polarized beams are available- since one
cannot gets any PV effect from QCD. Then, a non-zero PV asymmetry could
reveal
the presence of a new interaction. Of course, one has also to take
carefully into
account the influence of Standard electroweak interactions, which are well
controlled and give in any case a PV effect with a definite sign.

In a few years from now, the RHIC Spin Collaboration (RSC) will run the
heavy ion
collider RHIC in the $pp$ mode with polarized beams at a center-of-mass
energy
from 50 to 500 GeV \cite{RSC}. If one compares with the Tevatron, the
deficit in
energy is compensated by  the very high luminosity of the machine which
increases
with the energy, reaching ${\cal L} = 2. 10^{32} \, cm^{-2}.s^{-1}$ at 
$\sqrt s =
500$ GeV. These figures yield an integrated luminosity  \linebreak
$ \int{\cal L} dt
$ = 800 $pb^{-1}$ in a few months running.
\m
In a recent paper \cite{TV1} we have considered the inclusive production of
one high \ET jet at this machine. Focusing on the double
helicity PV asymmetry \ALLPV, we have found that the presence of a new PV
contact
interaction between quarks could yield some small but measurable
deviations from the expected SM asymmetry which is mainly due to the
interference
between QCD and Weak terms. The magnitude of the predicted asymmetry, in an
\ET range
between 60 GeV and 100 GeV,  allows to probe the compositeness scale up to
a value
$\Lambda \approx 2. - 2.5$ TeV, as soon as parity is maximally violated by
the new
effective interaction.

Concerning new massive vector bosons, many studies have considered the
Drell-Yan lepton pair production channel when the partonic c.m. is allowed
to be
of the same order of magnitude as the \ZP mass. In this case, relevant at
Hadronic Supercolliders only, it has been shown that polarized beams should
be
very efficient in the quest for the identification of the theoretical
origin of
the new boson \cite{FT12Casal}, a task which is much more difficult than
the
discovery itself (see e.g. Ref.\cite{CveticGodfrey}).

Given the maximal RHIC energy, if the \ZP is heavier than 800 GeV,
it is hopeless to get a signal in this channel and to measure a spin
asymmetry. 
Conversely, the hadronic channel is particularly well-suited for a \ZP
which goes
preferentially into quarks like the candidate from Refs.\cite{A,C}. 

In this paper, we assume that such a \ZP is indeed exchanged
in the $t$-channel of the quark-quark scattering process and 
we investigate the consequences  on the one-jet asymmetry \ALLPV which will
be
measured with great precision by the RSC.

\section{The parameter space} 

\m
One considers a new neutral vector boson which couples universally to $u$-
and 
$d$- type quarks with the following structure  : 
\EQ
{\cal L}_{Z'} = {g\over 2 \cos \theta_W} Z'^{\mu}{\bar q}_i \gamma_\mu[
C_{iL} (1 -
\gamma_5) \; +\; C_{iR} (1 + \gamma_5) ] q_i
\eq \no
for each given quark flavor $i$ (the \ZP ${\bar q}q$ couplings are
normalized
in the same way as the \Z ${\bar q}q$ ones). 
\m
Constraints from LEP/SLC data forces the leptonic couplings to be very
small. 
Neglecting
the latter, the \ZP width is given by the general  expression (for $M_{Z'}
>>
m_q$):   
\EQ
\label{largeur} 
\Gamma_{Z'} \; =\; N_g {G_F \, M_{Z}^2 \over {\pi \sqrt 2}} M_{Z'}
(C_{uL}^2 + C_{dL}^2 + C_{uR}^2 + C_{dR}^2)
\eq
\no
\m
The heavy \ZP which has been considered in Refs.\cite{A,C} does not belong
to any
particular model (there is a  vast literature on massive \ZP's,
corresponding to
various extensions of the SM, for a recent review one can consult Ref.
\cite{CveticGodfrey} and references therein). 
Since the analysis of
Ref. \cite{A} imposes more restrictions on the parameter space, we follow
these
authors first, and  for comparison we treat in a second step the extra
freedom
allowed in Ref. \cite{C}.
\m
In Ref. \cite{A}, a first constraint comes from assuming equality of the
left-handed couplings within one $SU(2)_L$ doublet : 
\EQ \label{leftequality}
C_{uL}\,= \, C_{dL}
\eq
On the other hand, the two right-handed couplings are left free. Following
the notations of Ref.\cite{A} one has $C_{uL}\,= \, C_{dL}\,  \equiv \,
x\,$;
$C_{uR} \, \equiv \, y_u $ and $C_{dR} \, \equiv \, y_d$.

\m
Then, after adjusting $M_{Z'}$ at 1 TeV, it seems possible to fit all the
LEP/SLC
observables, including $R_b$ and $R_c$, and the CDF jet cross sections in
the same
time.
 The quark couplings which are obtained are quite large. As a consequence
the \ZP
resonance can be very wide. We will consider as reasonable an upper limit
around 
500 GeV for $\Gamma_{Z'}$.
Therefore, we retain from Ref. \cite{A} the following values : 
$-1.5 \, \leq x \leq \, -0.5$, $2  \leq \, y_u \, \leq 2.5$ and $y_d \, =
\, 0$.
Note that the fit to LEP/SLC data is weakly sensitive to the last parameter
$y_d$. It will be also the case for our results presented below.
This parameter interval contains the "final fit" \cite{A} values 
for the parameters  : 
\EQ\label{ff}
x \, = \, -1, \; y_u\, = \, 2.2, \; y_d\, =\, 0
\eq
\m
Finally, using the approximate scaling
advocated in Ref. \cite{A} for the best fit values of the parameter :
\EQ \label{scaling}
x,y_u,y_d \; \sim \; {M_{Z'} \over M_Z}
\eq \no
one can easily control the (tiny) consequences of varying the \ZP mass down
to 
$\sim\, $ 800 GeV.
\m
In Ref.\cite{C} the authors have preferred to compare the magnitude of \ZP
couplings with those of the \Z coupling. They have defined the ratios of
the
vector and axial-vector couplings (with a
$\gamma_{\mu}[g_{Vi}-g_{Ai}\gamma_5]$
structure) : $\xi_{Vi(Ai)} \ \equiv \; g'_{Vi(Ai)}/g_{Vi(Ai)}$.
The relationship with the parameters $x,y_u,y_d$ is : 
\EQ \ST
\begin{array}{ccc} \ds
\xi_{Vi} \, = \, {x+y_i \over T_3^{i} - 2 Q^{i} \sin^2 \theta_W} & 
\; , \; & \ds
\xi_{Ai} \, = \, {x-y_i \over T_3^{i}}
 \end{array} \st \eq \no
where $T_3^{i}$ is the third component of the weak isospin of the quark $i$
and
$Q^{i}$ its electric charge.

Imposing some values of the couplings smaller than the QCD strength 
($\alpha_s \simeq 0.12$), one gets $|\xi_{Vu}| < 16 ; |\xi_{Vd}| < 10$ and
$|\xi_{Au,d}| < 7$.  The values we have retained above respect 
comfortably these constraints.  
In Ref. \cite{C} a smaller ($\leq 200$ GeV) \ZP width is preferred.
On the other hand, the constraint eq.( \ref{leftequality}) is not retained. 
Therefore, the parameter space is enlarged if we still impose  
$\Gamma_{Z'}\leq 500\,$GeV. Given this extra freedom, we choose two 
"extreme cases" which maximize PV effects, with a \ZP strongly coupled to
$u-$quarks, namely :

\no 
- a left-handed \ZP
\EQ\label{extremL}
\xi_{Au}=5.8 \; ; \; \xi_{Vu}=15.2 \; ; \; \xi_{Ad,Vd}= 0 \; {\rm
\;corresponding\;
to\;} \; x_u=2.9, y_u=0, x_d=y_d=0
\eq
\no
with $x_{u,d} \equiv C_{(u,d)L}$.

\no
- a right-handed \ZP is obtained by changing the sign of $\xi_{Au}\,$ :
\EQ\label{extremR}
\xi_{Au}=-5.8 \; ; \; \xi_{Vu}=15.2 \; ; \; \xi_{Ad,Vd}= 0 \;  
{\rm \;corresponding\; to\;}
\;  x_{u,d}=0, y_u=2.9, y_d=0
\eq
\no
In both cases we have $\Gamma_{Z'}\sim 500\,$GeV.
\m
The last parameter of these models is the amount of \Z - \ZP mixing. LEP
data
constrain in any cases the mixing angle $\xi_M$ to be very small (at most a
few
$10^{-3}$). Since we are interested into a direct effect of the \ZP we can
safely
neglect this mixing.  
\m

\section{Calculation and results}
It is easy to check that, at a variance with the situation at FNAL, at
RHIC,
due to the low value of the maximal c.m. energy , it is hopeless to get a
conclusive
signal from the measurement of the unpolarized $pp$ jet cross section
alone. We then
turn to the question of PV spin asymmetries. 
\m
To build a PV asymmetry one single polarized beam is
sufficient. However, our recent experience \cite{TV1}
taught us that a larger effect can be
obtained by using both polarized colliding beams which are available at
RHIC. One
can define the double helicity PV asymmetry :
\EQ
\label{ALLPVdef}
A_{LL}^{PV} ={d\sigma_{(-)(-)}-d\sigma_{(+)(+)}\over 
d\sigma_{(-)(-)}+d\sigma_{(+)(+)}}
\eq
\noindent
where the signs  $\pm$ refer to the helicities of the colliding protons.
The cross section $d\sigma_{(\lambda_1)(\lambda_2)}$ means the one-jet
production cross section in a given helicity configuration, \linebreak
$p_1^{(\lambda_1)}p_2^{(\lambda_2)} \r jet + X$, estimated at  $\sqrt{s} \
=\ 500$ GeV
for a given \ET, integrated over a pseudorapidity interval 
$\Delta \eta \, =\, 1$ centered at $\eta\,=\,0$. We have also integrated
over 
an \ET bin of 10 GeV. 
\m
In the \ET range of interest the contribution of quark-quark scattering
is by far dominant.  In short notations \ALLPV is given by the
expression : 
\EQ
\label{ALLPVjet}
A_{LL}^{PV} \simeq  {1\over d\sigma}\sum_{i,j} \sum_{\alpha,\beta}
\int
\left(T_{\alpha,\beta}^{--}(i,j) - T_{\alpha,\beta}^{++}(i,j)
\right) 
\biggl[q_i(x_1,\mu^2)\Delta q_j(x_2,\mu^2) + \Delta
q_i(x_1,\mu^2)q_j(x_2,\mu^2) 
+ (i\leftrightarrow j) \biggl]
\eq \no
The
$T_{\alpha,\beta}^{\lambda_1,\lambda_2}(i,j)$'s are the matrix element
squared
with  $\alpha$ boson and $\beta$ boson exchanges in a given helicity
configuration
for the involved partons $i$ and $j$. 
$\Delta q_i\ =\ q_{i+} - q_{i-}$ where 
$\, q_{i\pm}(x,\mu^2)$ are the distributions of 
the polarized quark of flavor $i$, either
with helicity parallel (+) or antiparallel (-) to the parent proton
helicity. Summing the two states one recovers  $q_{i+} + q_{i-} =
q_i(x,\mu^2)$.
All these distributions are evaluated at the scale $\mu\, =\, $\ET and we
have
checked that the precise choice for $\mu$ has no influence on our results.
The
unpolarized cross section $d\sigma$ is dominated by QCD and must also
include all the
relevant Electroweak+\ZP terms and their interference with QCD terms when
it is
allowed by color rules. Of course, the non-dominant $q(\bar q)g$ and $gg$
scattering
subprocesses have to be included in the part of the cross section which is
purely QCD.  
\m
Concerning the numerator of \ALLPV, the SM contribution has been already
discussed in
Ref.\cite{TV1}. The correct expressions for the relevant 
$T_{\alpha,\beta}$'s can be found in ref. \cite{BouGuiSof}.
Since we are in a regime where $\hat s << M_{Z'}^2$, the non-standard
effect will
be dominated by the interference between SM amplitudes and the amplitude
due to
\ZP exchange.  In fact,  90\% of the effect will come
from the interference between the \ZP and one-gluon exchange graphs for
identical
quarks. Due to color rules there is no such interference in case of quarks
of
different flavors. 

Then, for $q_iq_i \r q_iq_i$ one has:
\EQ \label{gZ'}
T_{g.Z'}^{\lambda_1,\lambda_2}(i,i) \, =\, {8 \over 9} \alpha_s\,\alpha_Z
\biggl[ (1 - \lambda_1)(1 - \lambda_2) C_{iL}^2 
+ (1 + \lambda_1)(1 + \lambda_2) C_{iR}^2 \biggl]\,
\hat s^2 Re \left( {1
\over \hat t D_{Z'}^{\hat u}} +
{1 \over \hat u D_{Z'}^{\hat t}}\right) 
\eq \no
where $\alpha_Z = \alpha/\sin^2\theta_W\cos^2\theta_W$,
and :
\EQ
D_{Z'}^{\hat t(\hat u)} \; =\; (\hat t(\hat u) - M_{Z'}^2) \, +\, 
i M_{Z'}\Gamma_{Z'}
\eq
Therefore: 
\EQ\label{simple}
 T_{g.Z'}^{--}(i,i) - T_{g.Z'}^{++}(i,i)
 \; \sim \; (C_{iL}^2 \, -\, C_{iR}^2)\, \times\, {\rm positive \; terms}
 \eq
\m
In the actual calculation, all the LO terms, involving quarks or
antiquarks, are added along with the very tiny Electroweak-\ZP interference
terms.
We restricted to LO terms since no NLO calculation for the \ZP contribution
is
available. \ALLPV being a ratio of cross-sections, we expect also that NLO
corrections
will not change the trend of our results.
\m 
Since real gluons are not involved in this
process, our results are not plagued by the uncertainties associated to
the imperfect knowledge of the polarized gluon distributions $\Delta
G(x,\mu^2)$. 
Concerning the polarized quark (and antiquark) densities, we have used some
recent
sets of distributions which fit all the available data.  The GS \cite{GS}
and GRV \cite{GRV} distributions yield very similar results for \ALLPV.
The effect obtained using the BS \cite{BS95} distributions being
simply larger in magnitude, we adopt a conservative attitude and we retain
the former
choice. It has to be kept in mind that our knowledge about these polarized
partonic
distributions will improve drastically in the future, thanks to the HERMES
experiment
at HERA \cite{HERMES} and, especially, thanks to the RSC program itself
\cite{BuncePw}. 

\m
We present in Fig. 1 the results of our complete calculation for \ALLPV
versus \ET,
including all terms, for $M_{Z'} = 1\,$ TeV (with GRV distributions). 
The expected SM asymmetry is shown for comparison. The error bars
correspond to the
statistical error for an integrated luminosity of $L_1 = 800 pb^{-1}$
(large bars) or
$L_2 = 4\times 800 pb^{-1}$ (smaller bars). As can be seen, the high
luminosity at
RHIC allows some very small statistical uncertainty \cite{BuncePw} on
\ALLPV. This
uncertainty is given roughly by : \EQ
\Delta A\; \simeq\; {1\over {\cal P}^2}\, {1\over
{\sqrt{N_{evts}^{++}\,+\,N_{evts}^{--}}}}
\eq \no
where ${\cal P}=0.7$ is the degree of polarization of one beam. As a
consequence, the
measurement of an asymmetry at the level of a percent or even less is
realistic in this
channel. 
\m
As already shown \cite{TV1,TannenbaumPenn}, the SM asymmetry due to
QCD-Weak 
interference is clearly positive : at \ET around 80 GeV, with $L_1$, it
lies at
2$\sigma$ above zero according to GRV or GS distributions (4$\sigma$ above
zero
according to BS distributions). The rise with \ET is due to the increasing
importance
of quark-quark scattering with respect to other subprocesses involving
gluons. 

We have checked that no "usual" \ZP from an extra U(1) or from a 
Left-Right symmetric extension of the SM can induce a sizable deviation
from the
expected SM  asymmetry. On the other hand, as can be seen from Fig. 1,
the presence of the hadrophilic \ZP introduces a strong perturbation.
With the "final fit" values of Ref.\cite{A} given in eq.(\ref{ff}), one
gets
for \ALLPV a value which is compatible with 0. It is clearly negative in
the most
favorable case corresponding to $x=-0.5, y_u=2.5, y_d=0$. For comparison,
we show the
consequences of the two extreme cases of eqs.(\ref{extremL},\ref{extremR})
corresponding to a Right-handed or a Left-handed \ZP allowed by the
analysis of
Ref.\cite{C}. 
\m
Of course, the statistical error also increases with
\ET at a fixed integrated luminosity. In Fig. 2, we show \ALLPV versus $x$ for
different $y_u$ values allowed in Ref.\cite{A} (with $y_d=0$), at a chosen
intermediate \ET = $80\pm 5$ GeV.  

Let us comment these results in more details :
\m
- First,  one has to note that the the $d-$quark polarized distribution
$\Delta d$ (which is usually negative) has a minor influence on our results
: as a
consequence moving the right-handed coupling $y_d$ away from the zero value
has no
visible effect on \ALLPV. Note also that changing the signs of the left or
right couplings has no influence on \ALLPV (see eq.( \ref{simple})).

- The behavior of \ALLPV is then governed by the ratio $|x|/y_u$ : \ALLPV 
decreases and tends to become negative when this ratio decreases.  
Let us consider the restricted parameter space of Ref.\cite{A} according to
Section 2.  Taking an integrated luminosity $L_1$, one gets an asymmetry
which is at least more than 1$\sigma$ below its SM value, provide that
$|x|<$1, and
2$\sigma$ below in the most favorable situation of large $y_u$ and small
$|x|$. In any
case, the effect should be confirmed without any ambiguity with the
enlarged statistics
corresponding to $L_2$.
\m
- If one applies the scaling according to eq.(\ref{scaling}), 
lowering the \ZP mass down to 800 GeV doesn't change appreciably the values
of
\ALLPV since the quark couplings are lowered in the same time.
\m
- The "extreme" cases from Ref.\cite{C} are also shown on this plot, along
with the
very similar results one gets from the presence of an effective handed
interaction
due to a contact term between quarks \cite{EichtenEHLQ} (calculations are
described in
Ref.\cite{TV1}) : 
\EQ\label{Lcontact}
{\cal L}_{qqqq} = \epsilon \, {\pi\over {2 \Lambda_{qqqq}^2}} 
\, \bar \Psi \gamma_\mu (1 - \eta \gamma_5) \Psi . \bar \Psi
\gamma^\mu (1 - \eta \gamma_5) \Psi
\eq
\noindent
where $\Psi$ is a quark doublet and $\epsilon$ a sign. 
 We have chosen the case of constructive interference
with QCD amplitudes ($\epsilon = -1$, see \cite{EichtenEHLQ,TV1}) and of
maximal PV
corresponding to  $\eta=1$(left-handed) or -1 (right-handed), with the
$\Lambda_{qqqq}$ value extracted by CDF. 
\linebreak
In these cases the effects on \ALLPV are
important, which is not surprising since $\Lambda_{qqqq} \sim 1.6\,$TeV is
a
relatively low value according to our previous calculations \cite{TV1}.

Of course, the distinction between
a possible \ZP effect and an  effect due to compositeness is quite
artificial
since it relies on the particular normalization of the contact interaction
and on the
precise chosen value for $\eta$ in eq.(\ref{Lcontact}). Only experiments at
higher
energies and/or a careful study of the invariant mass distribution of
di-jets events
at the Tevatron should allow to disentangle the two possibilities.

\section{Conclusion} 

In this letter we have shown that a new \ZP, with a mass in the TeV range,
could induce
some visible effects in the asymmetry \ALLPV for inclusive one-jet
production in the
collision of polarized beams at $\sqrt s = 500$ GeV with a high luminosity.

The two conditions which are to be fulfilled are an enhanced coupling of
the \ZP to
quarks and an important asymmetry between Left and Right-handed couplings.
The expected SM asymmetry being small and positive, the most spectacular
effect
should come from the observation of a zero or negative asymmetry or of a
large and
positive one. 
According to Ref.\cite{A}, $y_u  \;{ ^>_{\sim}} \,2 \,|x|$ is a general
tendency
of the fitted values of the parameters. It means that, concerning
$u-$quarks, 
there is a dominance of
the right coupling on the left coupling to the \ZP. 
Furthermore, the role of $d-$quarks, which is not dominant, is further
suppressed by the choice $y_d=0$. Then, the predicted values of \ALLPV tend
to be
significantly below the positive values expected in the SM. 
On the other hand, the parameter space of Ref.\cite{C} is larger, which
implies more
freedom for the resulting \ALLPV values. However, pure Left-handed as well
as pure
Right-handed interactions are allowed : these two cases give of course some
large
effects.

\m
It can be surprising that measuring a spin asymmetry gives more chances for a
discovery than the simpler measurement of an unpolarized cross section. In
fact, the
situation is reminiscent of the polarized DIS experiment performed at SLAC
in the
seventies \cite{Prescott} where an effect of $\sim \, 10^{-4}$, due to the
Standard
\Z, was seen in the PV left-right helicity asymmetry. 
In the same time, such an electroweak effect was far too small to be
observed from the
unpolarized cross section. 

The availability of high-intensity polarized beams at RHIC should allow to
carry on the
same kind of program, in the more difficult context of hadronic collisions.

\vspace*{3cm}

\no {\bf Acknowledgments}

\m \no
We thank P. Chiappetta and C. Verzegnassi for discussions about
the hadrophilic \ZP, and 
M. Tannenbaum for encouraging us to investigate Non Standards effects at
RHIC.


%
\newpage

\vspace*{4cm}
{\bf Figure captions}
\bigbreak
\no
{\bf Fig. 1} \ALLPV for one-jet inclusive production, versus \ET, at 
$\sqrt s = 500$ GeV, integrated luminosities $L_1$ (large error bars) and
$L_2$ (small error bars). SM expectation (plain curve); \ZP effect with
"final fit"
values $x=-1,\,y_u=2.2,\,y_d=0$ (dotted curve), maximal effect with
$x=-0.5,\,y_u=2.5,\,y_d=0$ (dashed curve) both from Ref.\cite{A} ; 
$Z'_{L}max$ and $Z'_{R}max$ correspond to two extreme cases from
Ref.\cite{C}
with maximal parity violation.
The calculations are performed with GRV distributions.
\bigbreak
\no
{\bf Fig. 2} \ALLPV at $E_T\,$= 80 GeV versus $x$ with different values for
$y_u$ and $y_d=0$ (same conditions as in Fig.1). 
The dotted lines correspond to $y_u \, =\,$ 2.0; 2.2; 2.5 from top to
bottom.
Also shown the two cases  $\eta = \pm 1$ for a compositeness scale
$\Lambda = 1.6\,$ TeV and the positions of the extreme cases  
$Z'_{L}max$ and $Z'_{R}max$.

\bigbreak
\no

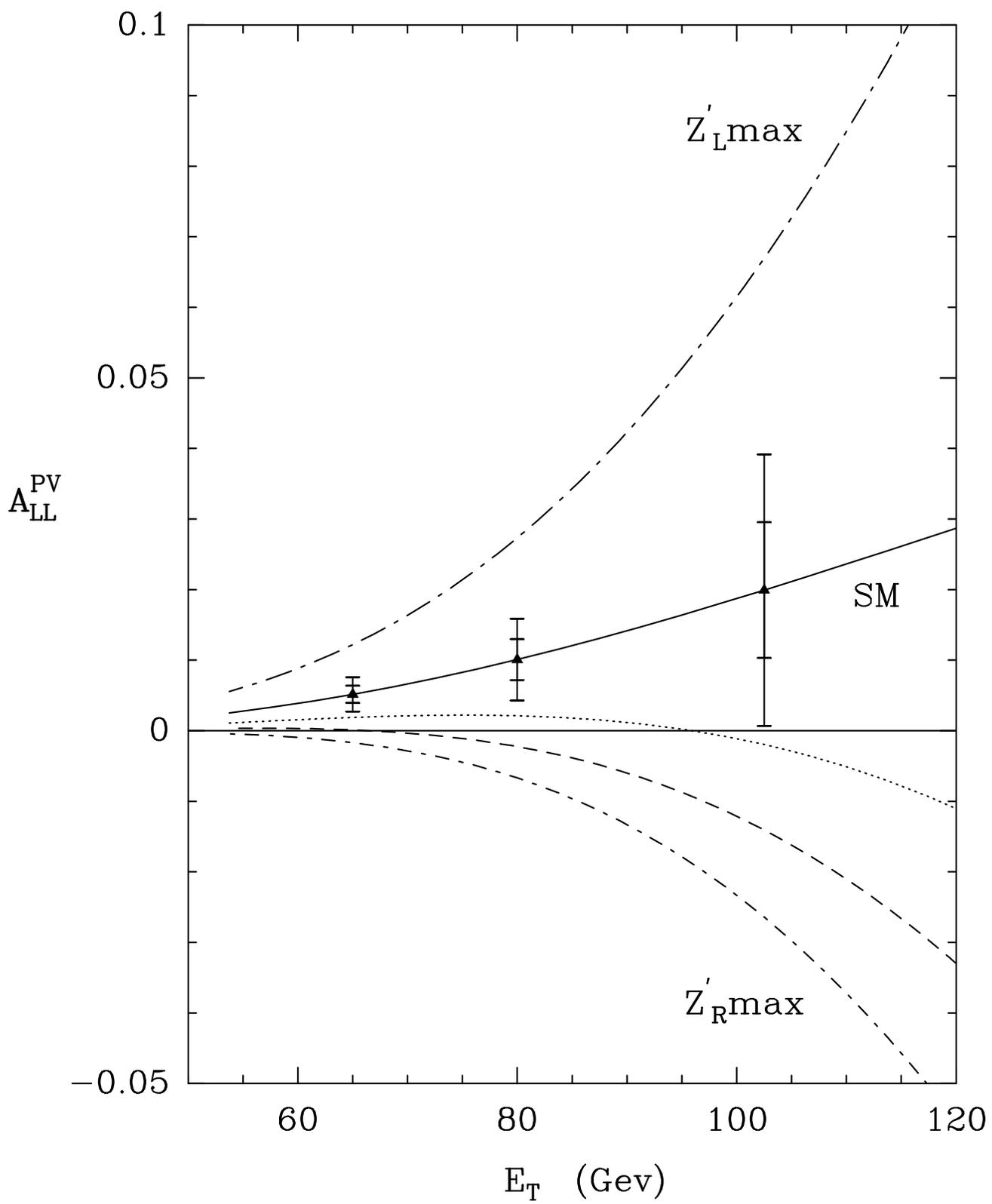
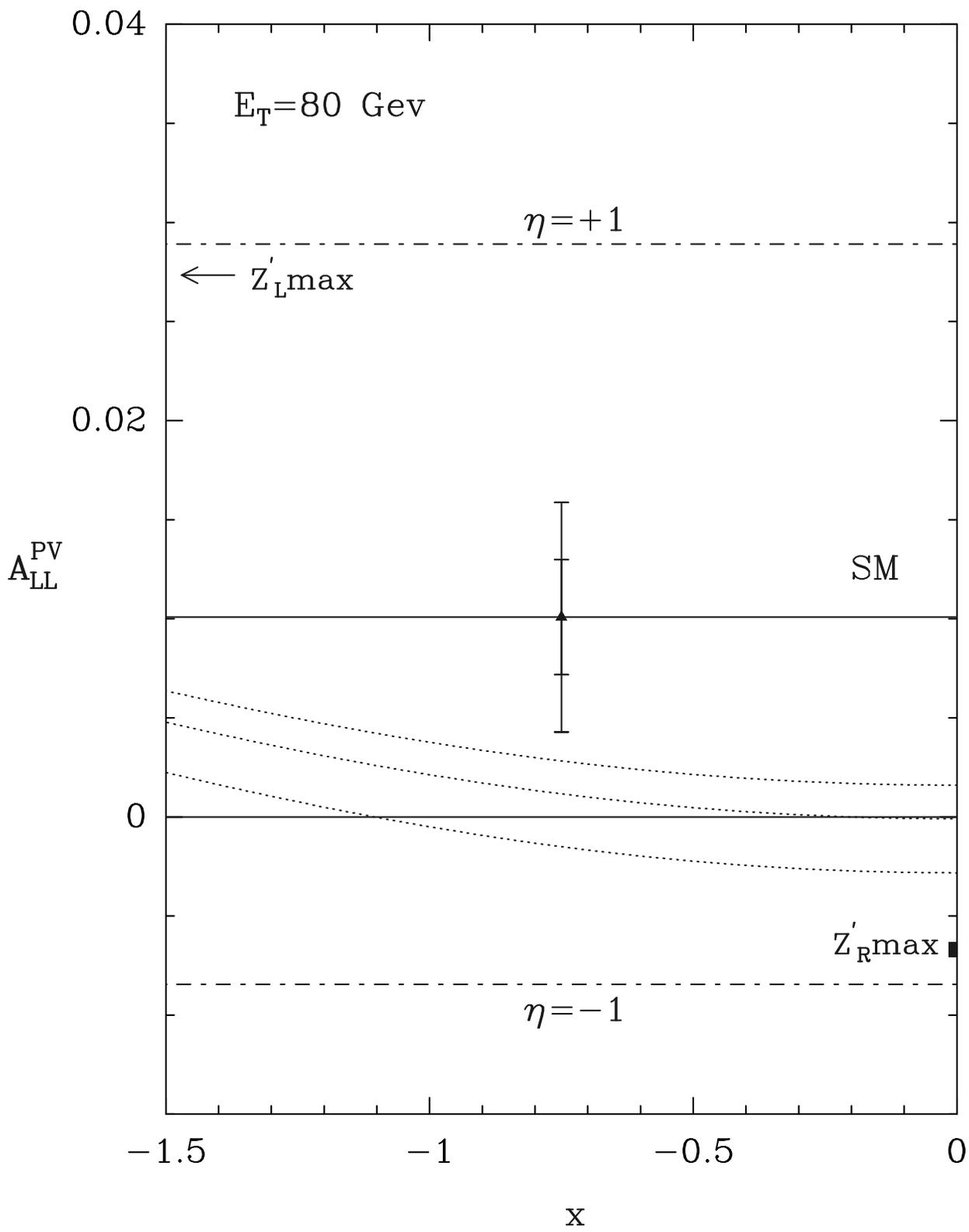
\end{document}